\documentclass[conference]{IEEEtran}
\IEEEoverridecommandlockouts
\usepackage{cite}
\usepackage{soul}
\usepackage{amsmath,amssymb,amsfonts}
\usepackage{algorithmic}
\usepackage{graphicx}
\usepackage{textcomp}
\usepackage{caption}
\usepackage{xcolor}
\usepackage{subcaption}
\def\BibTeX{{\rm B\kern-.05em{\sc i\kern-.025em b}\kern-.08em
    T\kern-.1667em\lower.7ex\hbox{E}\kern-.125emX}}
    \def\undb#1{\mbox{\bf{#1}}}

\begin{document}

\title{Optimizing a Digital Twin for Fault Diagnosis in Grid Connected Inverters -- A Bayesian Approach\\
%\thanks{This work is partly supported by the FIREMAN project
%CHIST-ERA-17-BDSI-003 funded by the Spanish National
%Foundation (PCI2019-103780) and the Academy of Finland (AoF;
%n.326270). This work is also partly supported by AoF: %EnergyNet n.321265/n.328869 and by JAES Finnish Foundation
%via STREAM project.}
}

\makeatletter
\newcommand{\linebreakand}{%
  \end{@IEEEauthorhalign}
  \hfill\mbox{}\par
  \mbox{}\hfill\begin{@IEEEauthorhalign}
}
\makeatother

\author{\IEEEauthorblockN{Pavol Mulinka}
\IEEEauthorblockA{\textit{Sustainable Artificial Intelligence Research Unit} \\
\textit{Centre Tecnol\`{o}gic de Telecomunicacions de Catalunya (CTTC/CERCA)}\\
Barcelona, Spain \\
pmulinka@cttc.es}
\and
\IEEEauthorblockN{Subham Sahoo}
\IEEEauthorblockA{\textit{Department of Energy} \\
\textit{Aalborg University}\\
Aalborg East, Denmark \\
sssa@energy.aau.dk}
\linebreakand
\IEEEauthorblockN{Charalampos Kalalas}
\IEEEauthorblockA{\textit{Sustainable Artificial Intelligence Research Unit} \\
\textit{Centre Tecnol\`{o}gic de Telecomunicacions de Catalunya (CTTC/CERCA)}\\
Barcelona, Spain \\
ckalalas@cttc.es}
\and
\IEEEauthorblockN{Pedro H. J. Nardelli}
\IEEEauthorblockA{\textit{School of Energy Systems} \\
\textit{LUT University}\\
Lappeenranta, Finland \\
pedro.nardelli@lut.fi}
}

\maketitle

\begin{abstract}
In this paper, a hyperparameter tuning based Bayesian optimization of digital twins is carried out to diagnose various faults in grid connected inverters. As fault detection and diagnosis require very high precision, we channelize our efforts towards an online optimization of the digital twins, which, in turn, allows a flexible implementation with limited amount of data. As a result, the proposed framework not only becomes a practical solution for model versioning and deployment of digital twins design with limited data, but also allows integration of deep learning tools to improve the hyperparameter tuning capabilities. For classification performance assessment, we consider different fault cases in virtual synchronous generator (VSG) controlled grid-forming converters and demonstrate the efficacy of our approach. Our research outcomes reveal the increased accuracy and fidelity levels achieved by our digital twin design, overcoming the shortcomings of traditional hyperparameter tuning methods. 
\end{abstract}


\section{Introduction}\label{sec:intro}
The concept of cyber-physical system 
%as a specific object constituted by 
comprising physical, data and decision layers has become an important research topic across different domains \cite{nardelli2022cyber}.
An essential aspect of such systems is the quantification of attributes that enable data-driven approaches for improved operation.
In particular, automated anomaly detection and diagnostics are data-driven tasks that would benefit from novel statistical methods.

In this paper, our focus is on fault diagnosis in grid connected inverters.
Although many efforts have been carried out on virtual synchronous generators (VSGs) to expedite stability under different grid conditions, their operation under faults or large signal disturbances still remains a challenge \cite{rocabert2012control}. Anomaly detection strategies are thus required to enhance  security and reliability, and enable  a widespread penetration of converters in the grid  \cite{he2020}. 
Most of the anomaly detection methods in the literature rely on classic strategies like switching pattern and voltage observers \cite{bec2021, zhou2018} or frequency analysis \cite{lezana2006}.
Despite the effectiveness of these approaches, they are quite application dependent, focused mainly on the modulation techniques. To this end, digital twins have recently gained popularity as an emerging digital technology for monitoring long-term and short-term anomalies \cite{8984243}, \cite{9141430}. 

Digital twins, as a concept associated with cyber-physical integration, constitute virtual representations of physical assets used for condition monitoring, fault detection, predictive diagnostics, improved maintenance regiments and reduced downtime. 
Their prototypical hallmark lies on the ability to harness tools from artificial intelligence, machine learning and advanced analytics in order to ingest and transform enormous quantities of production data into actionable insights.
Although digital twin adoption is capable of revolutionizing the operation of physical components, numerous challenges exist, mainly related to their scalable and resilient design, high-fidelity modularity, and synchronization between digital and physical counterparts.
In addition, the efficiency of digital twins is largely administered by the quantity as well as quality of data \cite{9595231}. 

To address such challenges in the context of power electronic systems, we apply online optimization on a digital twin of a VSG controlled grid-forming converter using a Bayesian optimization framework.
Since the proper configuration of hyperparameters (e.g., regularization weights, learning rates) in machine learning algorithms is crucial for successful generalization, we employ a computationally efficient methodology, which tunes the hyperparameters with formal guarantees that the uncertainty can be reduced over time. 
To demonstrate the  proposed optimization framework, we use the following dynamic model of the VSG controlled grid-forming converters \cite{vsg}, as illustrated in Fig. \ref{fig:sys}:
\begin{eqnarray}
\dot{\undb{x}} = \undb{A}\undb{x} + \undb{B}\undb{u}
\end{eqnarray}
where, $\undb{x} = [v_g, i_o, \omega, v_{dc}, P, Q], \ \undb{u}=[P_{ref}, Q_{ref}]$ for the anomalies: (i) line-to-line faults, (ii) sensor faults, (iii) single-phase voltage sag and (iv) three-phase faults. Hence, we treat the fault diagnosis problem as a multi-class classification task. Our model-based approach defines a prior belief over the possible objective functions, and then sequentially refines the model as data are observed (i.e., function evaluations) via Bayesian posterior updating. 
We demonstrate the efficiency of our framework considering different fault cases with limited amount of training data to explore hyperparameter combinations. Performance evaluation reveals that
our method is able to exploit regions of the hyperparameter space that are already known to be promising, avoiding redundant and costly function evaluations that may otherwise render hyperparameter tuning a cumbersome and time-consuming process.
%\begin{table}
%\centering
%\caption{Fault description}
%\begin{tabular}{|l|c|l|}
%    \hline
%	\textbf{Class} & \textbf{Label} & \textbf{Description} \\ \hline
%    0 & Normal behavior &  \\ \hline
%	1 & LL\_Fault & TBD\_SUBHAM \\ \hline
%	2 & Three-Phase\_Sensor\_Fault & TBD\_SUBHAM \\ \hline
%	3 & Single\_Phase\_Sag & TBD\_SUBHAM \\ \hline
%	4 & Three\_Phase\_Grid\_Fault & TBD\_SUBHAM \\ \hline
%\end{tabular}%
%\label{fault_desc}
%\caption{Fault description.}
%\end{table}

\begin{figure}[t!]
	\centering
		\includegraphics[width=\columnwidth]{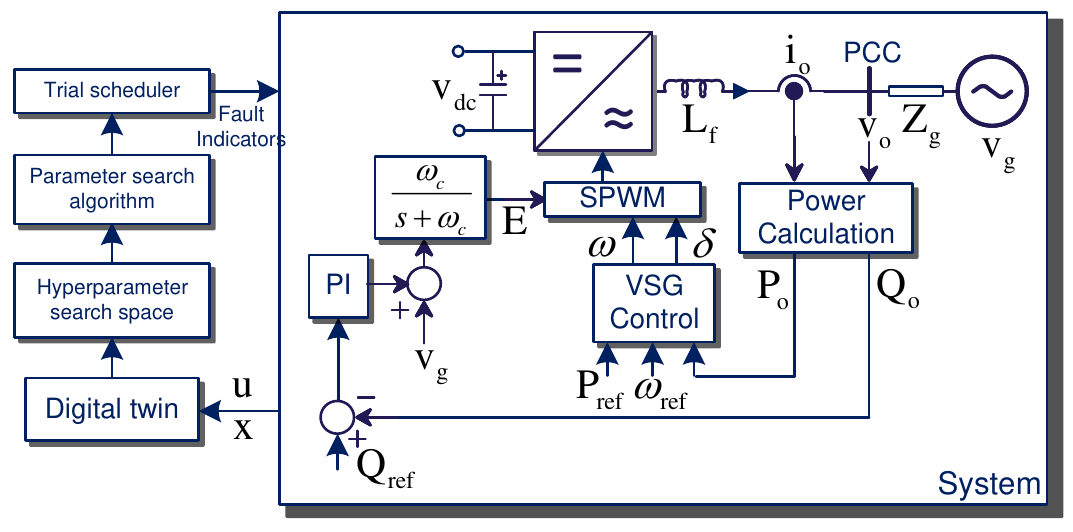}
    \captionsetup{font=footnotesize}
	\caption{Optimized digital twin framework for a VSG controlled grid-forming converter.}
	\label{fig:sys}
\end{figure}%

\section{Problem Statement}\label{sec:soa}
Installation of devices such as active power filter (APF) or developing new control strategies of grid-connected inverters with the capability of mitigation of disturbances is the case of attention for the research community, and industries. However, the APF has a limited amount of power storage to mitigate long-lasting power fluctuations and voltage dips in medium and high voltage levels. Furthermore, they are additional devices and add cost to electric power systems. The main drawback with grid-forming inverters is that in the case
of sharp changes in the amount of load or faults that has a high $dv/dt$, $df/dt$, the threshold defined for the changes of voltage/frequency in the power generation control and their protection system leads to unnecessary load shedding and the
tripping of generation \cite{nsoni}. Lately, grid forming inverters are getting more popular to overcome aforementioned drawbacks \cite{vst,hl}. However, its response to short circuit and other contingencies still remain in question, which can easily lead to loss of synchronization. 

To detect these anomalies and direct them to equipped countermeasures for power electronic systems, many solutions have been devised in the literature. In \cite{kaus}, a data-driven mechanism is used to characterize between different physical anomalies in power electronic systems. However, in distribution systems, although information exchange through cyber layer facilitates the operation of power electronics, it also exposes the system to cyber-attacks \cite{csec}. In this regard, many anomaly detection solutions have been proposed in \cite{kg}, \cite{mak}. As cyber-attacks can be augmented using a smart policy and can be replicated very close to that of grid faults, a physics-informed spline learning strategy is also proposed in \cite{pisl} to provide anomalous detection with minimal data. Another categorical approach could be to clone a digital twin of the system to determine anomalous behavior. However, digital twin can be ill governed based on perturbed data, which requires optimization. Hence, in this paper, we propose an optimization framework using Bayesian framework to tune the digital twin based on critical hyperparameters.

\section{Proposed Optimization Framework of Digital Twin}\label{sec:two}
Our goal is to design a surrogate single-input single-output (SISO) model \cite{huang2021simplified} of the considered power electronic system, by broadening its domain knowledge using steady-state features as well as fault instances. To this end, we consider a relatively small dataset of $100011$ set-points to expedite the learning capabilities of digital twin using the proposed optimization framework. We then split the dataset into training, validation and testing parts with an $8:1:1$ ratio, following the machine learning performance evaluation practices. In testing, we combine validation and testing datasets, since both of them were not used for training of the models. The solution for the grid-forming converter use case is then adjusted to implement an experiment tracking, model versioning and deployment block as part of the pre-existing solution, illustrated in Fig.~\ref{fig:syst}. 

\begin{figure}[t!]
\centering
		\includegraphics[width=\columnwidth]{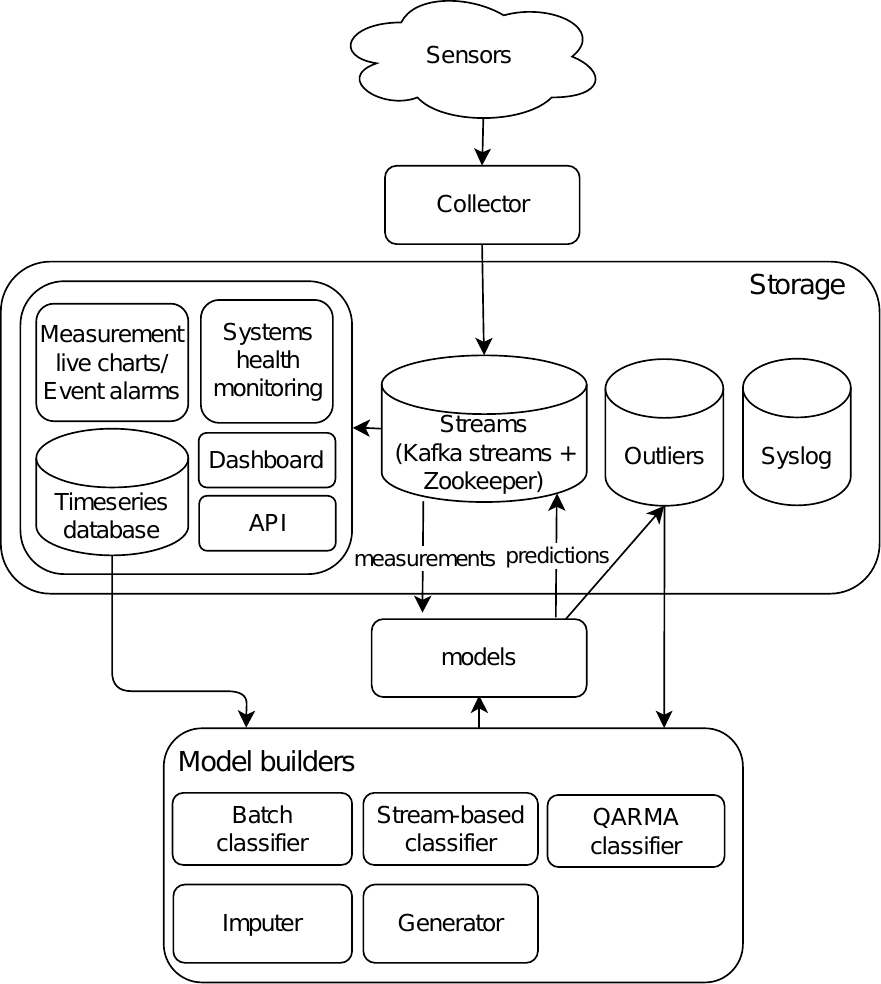}
    \captionsetup{font=footnotesize}
	\caption{Schematic of the proposed optimization framework of digital twin, with pre-existing and publicly available modules for fault detection and classification \cite{mulinka2021information}. In this paper, we focus on improving the \emph{models} and \emph{Model builders} blocks.}
	\label{fig:syst}
\end{figure}

\subsection{Model selection}

We follow a structured approach to design a solution from \textit{out-of-the-box} open-source components, and combine them to obtain the best possible evaluation outcomes. Our focus is on the availability of the state-of-the-art methods, with a potential to use the solution in the production environment, by leveraging scheduling, parallelism, early-stopping and model deployment of components, i.e., provide production-ready solution based on most recent machine learning research and development. In particular, we narrow down the search for the best suitable model for the digital twin into four categories:
\begin{itemize}
  \item \textbf{AutoML baseline model}: We use automatic machine learning (AutoML) implementation by H2O~\cite{H2OAutoML20} as a baseline for our adjusted and fine-tuned models.
  \item \textbf{Traditional baseline models}: We use traditional machine learning methods to supplement complex ensemble methods from AutoML.
  \item \textbf{Gradient boosting machines (GBM) models}: GBMs are one of the most popular methods used to solve a variety of tasks in Kaggle competitions \cite{Kaggle}. In this work, we focus on LightGBM due to its performance and easy-to-use application programming interface (API) to implement custom objective and loss functions~\cite{ke2017lightgbm}. 
  \item \textbf{Deep learning models}: We use state-of-the-art machine learning models, including  Tabtransformer~\cite{huang2020tabtransformer} and Tabnet~\cite{arik2020tabnet}.
\end{itemize}

In both LightGBM and deep learning models, we leverage a custom implementation of focal loss function for the case of multi-class classification of imbalanced tabular datasets~\cite{lin2017focal}. To orchestrate our experiments, we use a Ray Tune cluster
%~\cite{liaw2018tune} cluster as out-of-the-box it 
including the Bayesian parameter search algorithm
%~\cite{bayes_opt} 
for optimization of algorithm hyperparameters and the asynchronous successive halving algorithm (ASHA) scheduler
%~\cite{li2020system} 
to accommodate early stopping of the bad performing trials. Experiment tracking, model versioning and deployment are performed using the Weights and Biases platform under academic license~\cite{wandb}.

\subsection{Model parameter optimization}

Our optimization framework introduces a reconfigurable and generally applicable approach that we make publicly available on the FIREMAN project GitHub repository \cite{pmulinkagithub}. In general, the machine learning pipeline preprocesses the dataset and incrementally drops features, i.e., measurements, to improve model performance. This iterative process is illustrated in Fig.~\ref{fig:tuning_cycle}. More specifically: 

\begin{enumerate}
    \item Parameters of the model are tuned using RayTune~\cite{liaw2018tune};
    \item Performance metrics (e.g., precision, recall, F1-score) of the best trained model are stored;
    \item Features are sorted according to their importance (i.e., inherent attribute of GBMs) or SHAP values~\cite{NIPS2017_7062}, and the feature with the worst achieved performance is dropped from the dataset.
\end{enumerate}

The iterative process of tuning and feature dropping is repeated until the model performance is not improving anymore. We consider this as an adjusted notion of \emph{early stopping}, a well known regularization technique, for parameter and model tuning. Instead of stopping model training 
%not continuing to train the model 
when the performance is not further improved, we incrementally drop features and retrain/re-tune the model until performance gains become saturated.
%is not further improved.
%

\begin{figure}[t!]
\centering
    \begin{subfigure}{\columnwidth}
        \centering
        \includegraphics[width=.5\columnwidth]{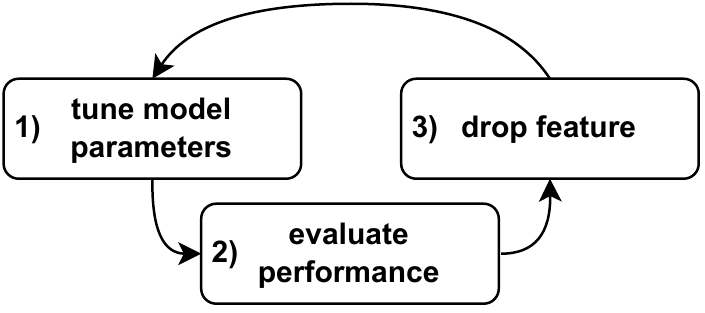}
        \caption{\footnotesize Feature dropping and parameter tuning iterative process. Features are incrementally dropped until the model performance is not further improved.}
        \label{fig:tuning_cycle}
    \end{subfigure}

\vspace{0.33cm}
\begin{subfigure}{\columnwidth}
        \centering
        \includegraphics[width=\columnwidth]{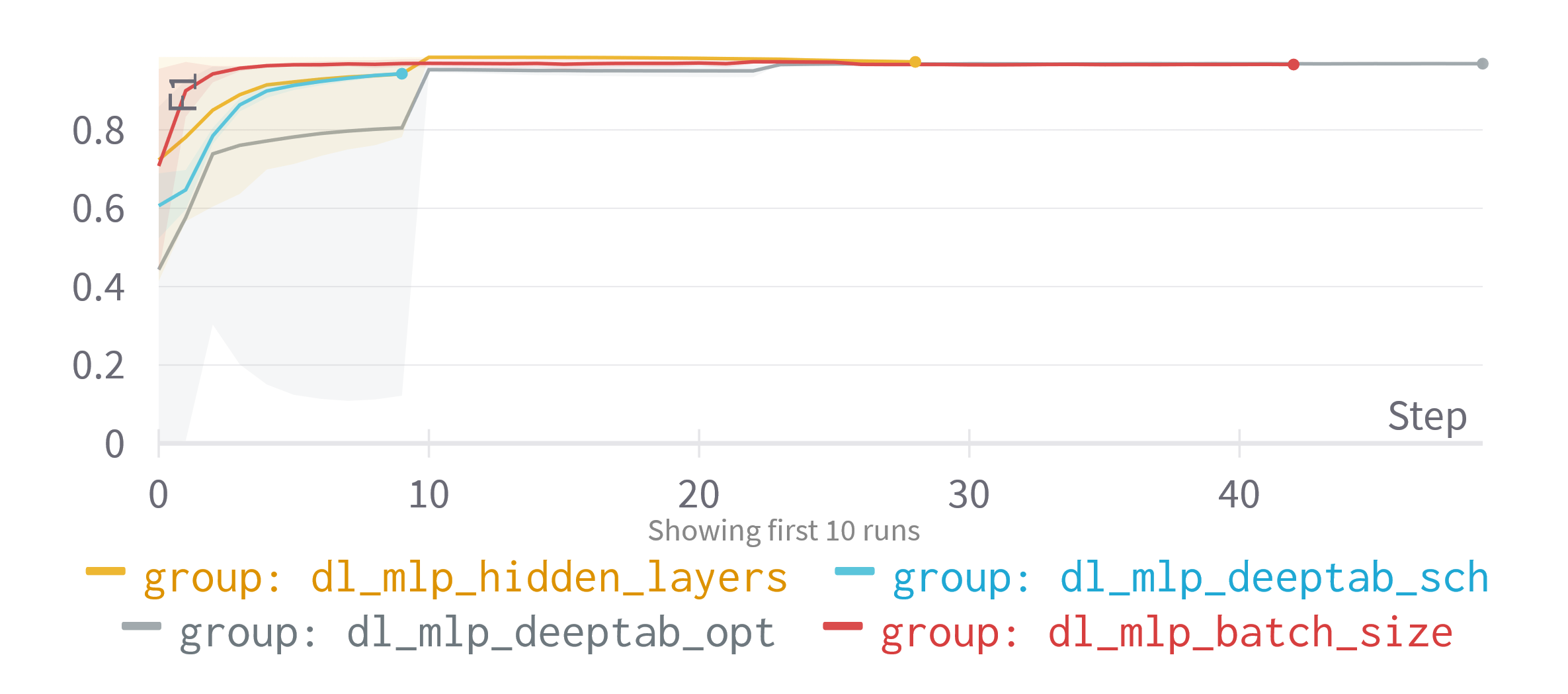}
        \caption{\footnotesize Step-wise parameter tuning of a multilayer perceptron network. Deep learning parameters tuned in a \textit{step-wise} manner are: hidden layers, scheduler, optimizer and batch size. Performance evaluation metric is F1 score (y-axis). Color-filled areas below the curves show the performance range of the parameter values within the tuning group, from min. to max.}
        \label{fig:dl_step_wise_tuning}
    \end{subfigure}%
%\vspace{0.3cm}
    \captionsetup{font=footnotesize}
	\caption{Model tuning procedure.}
	\label{fig:tuning}
\end{figure}

We also perform parameter tuning of the deep learning models under consideration. The task is inherently more complex than tuning traditional machine learning models. Deep learning models include not only numerical parameters, e.g., batch size and learning rate, but also choice of activation, objective and optimization functions and design of the layers, to name a few. To address such complexity, we follow a \textit{define-by-run} approach used by authors of Optuna~\cite{optuna_2019} to adjust model parameters during the training. In particular, 
%more specifically their 
a \emph{step-wise} method implemented in LightGBM tuner is followed, where independent sampling of the parameters is performed, i.e., sampling parameter search space for each parameter with fixed values of other parameters. This approach can thus be summarized in three steps: 

\begin{enumerate}
    \item Define value search space for every parameter;
    \item Define order of the parameters;
    \item Iterate over the parameters and search for best performing values.
\end{enumerate}

In the case of deep networks, the value search space can be a set of optimization functions and their parameter values, e.g., stochastic gradient descent (SGD) and ADAM optimizers and their learning rates, number and size of the hidden layers or training batch size. Each step is referenced as a group of parameter values. Fig.~\ref{fig:dl_step_wise_tuning} shows the process of step-wise tuning of hidden layers, scheduler, optimizer and batch size, for a multilayer perceptron network. %Order of the tuning. 
In each step, a model parameter is tuned, setting fixed values for the other parameters. After each step, the tuned parameter is saved and used in the next step. It is noted that the resulting parameter value set depends on the tuning order and may not correspond to the best parameter combination. However, to our knowledge, it is currently the best alternative to exhaustive parameter grid search.

\begin{table}[t!]
\centering
\captionsetup{font=footnotesize}
\caption{Class encoding.}
\begin{tabular}{|l|l|}
            \hline
        	\textbf{Class \#} & \textbf{Description}\\ \hline
                    0 & Normal behavior \\ \hline
                    1 & Line-to-line fault \\ \hline
                    2 & Three-phase sensor fault \\ \hline
                    3 & Single-phase sag \\ \hline
                    4 & Three-phase grid fault \\ \hline
        \end{tabular}
        \label{tab:desc}
\end{table}

\subsection{Output explanation}
Fig.~\ref{fig:explanation} illustrates a per fault algorithm prediction explanation for the best performing model, LightGBM. Decision plots show how each feature (i.e., measurement) contributes to each type of fault and normal behavior prediction, and a final prediction probability for each class. Class number to fault type encoding is summarized in Table~\ref{tab:desc}. It can be observed that our approach poses elevated merit
%We see the highest added value of this approach 
in complex multi-dimensional scenarios, where it is not clear which is the root cause of the true or false fault prediction of the algorithm.

\begin{figure}
\centering
    \begin{subfigure}{0.4725\linewidth}
        \includegraphics[width=\linewidth]{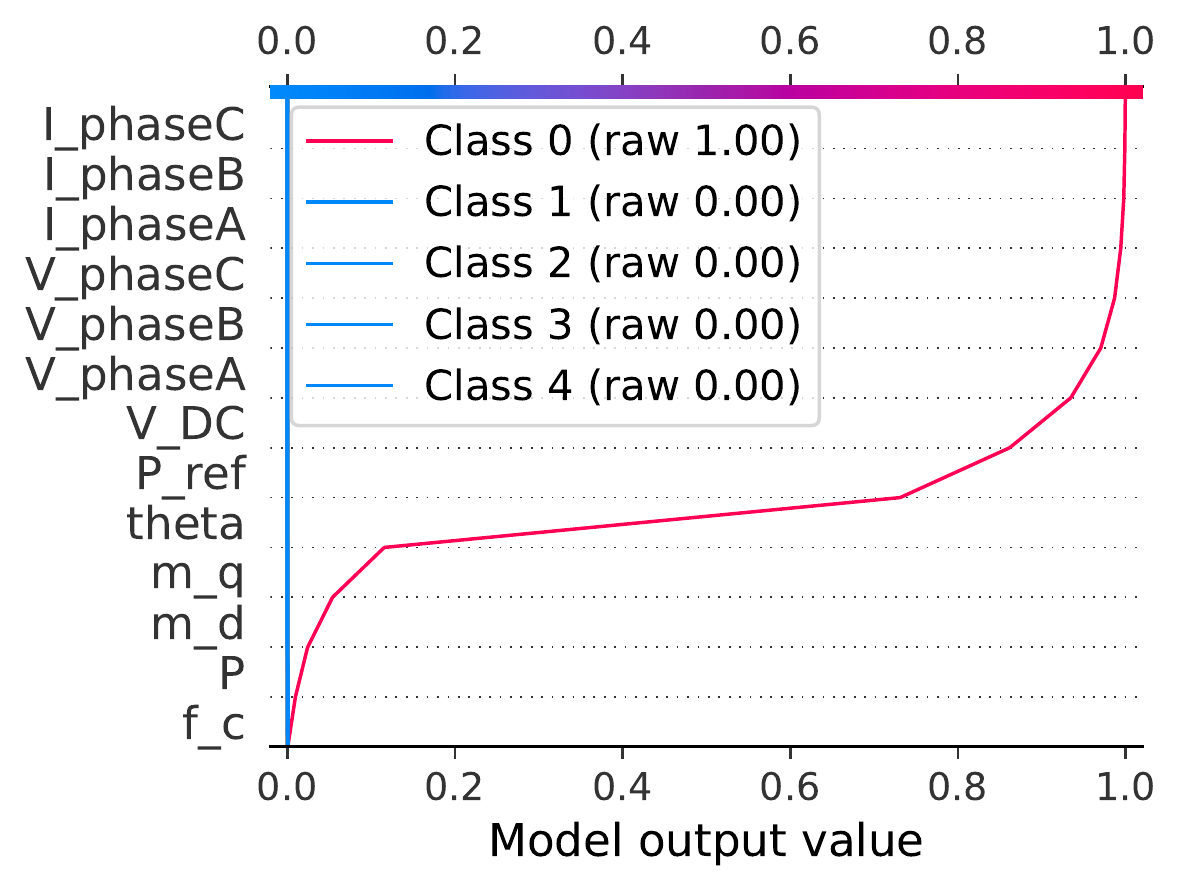}
        \caption{Normal behavior}
        \label{fig:normal_shap}
    \end{subfigure}    

\vspace{0.3cm}
    \begin{subfigure}{0.4725\linewidth}
        \includegraphics[width=\linewidth]{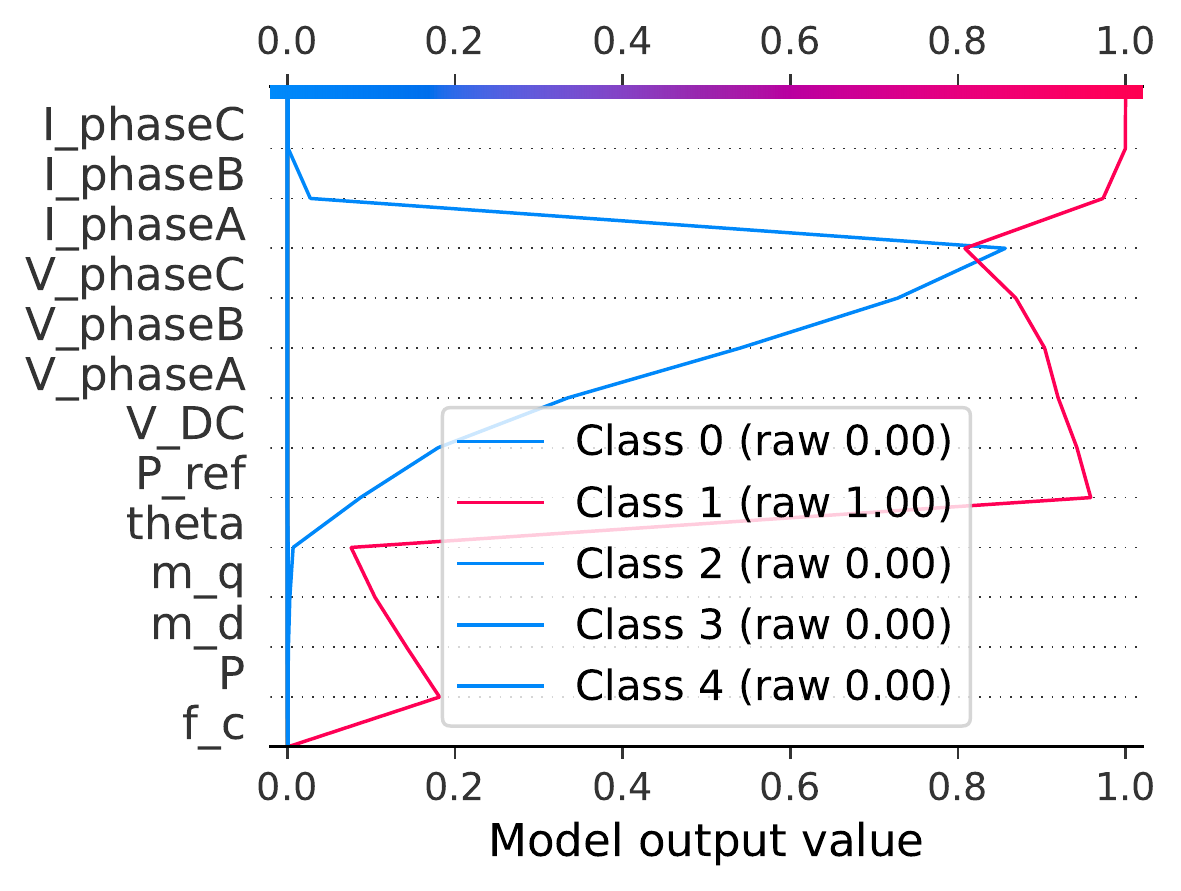}
        \caption{Line-to-line}
        \label{fig:LL_Fault_shap}
    \end{subfigure}%
\vspace{0.3cm}
    \begin{subfigure}{0.4725\linewidth}
        \includegraphics[width=\linewidth]{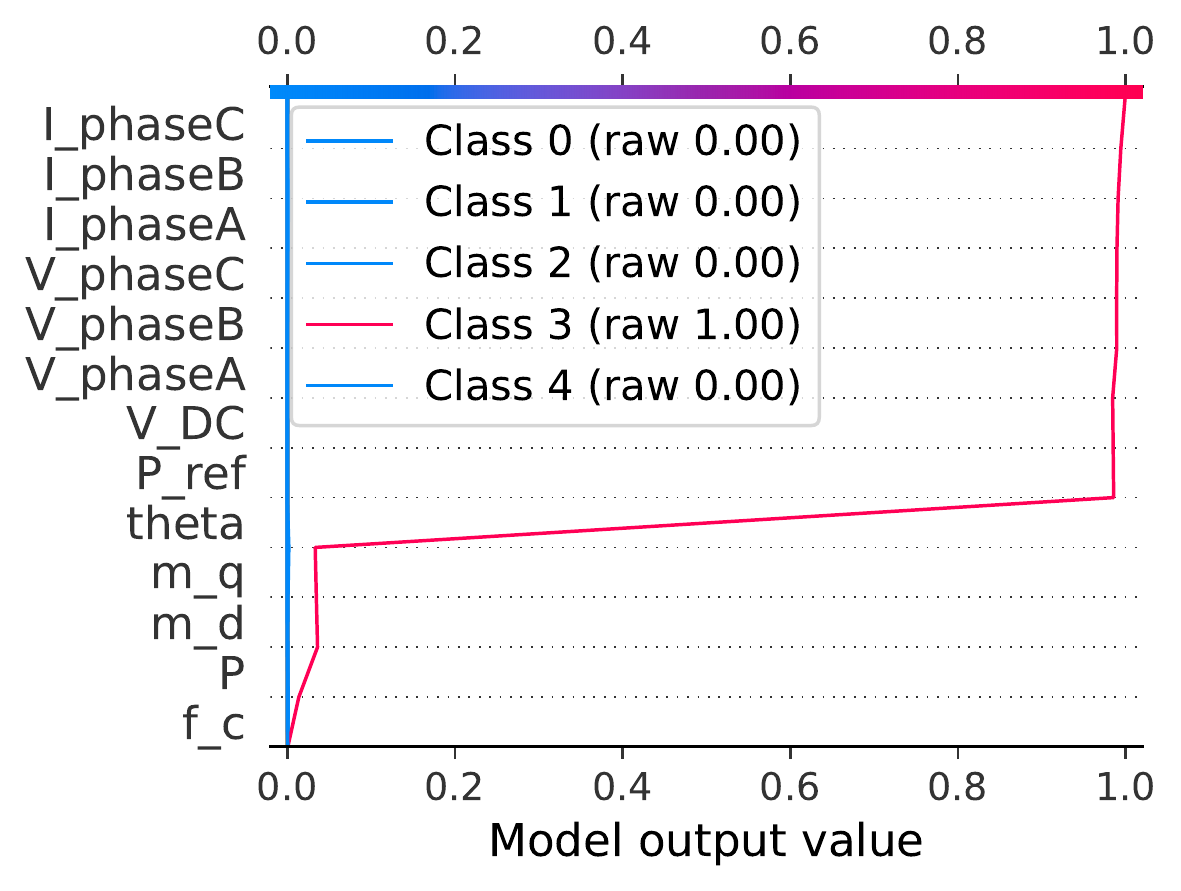}
        \caption{Single-phase sag}
        \label{fig:Single_Phase_Sag_shap}
    \end{subfigure}%

\vspace{0.3cm}
    \begin{subfigure}{0.4725\linewidth}
        \includegraphics[width=\linewidth]{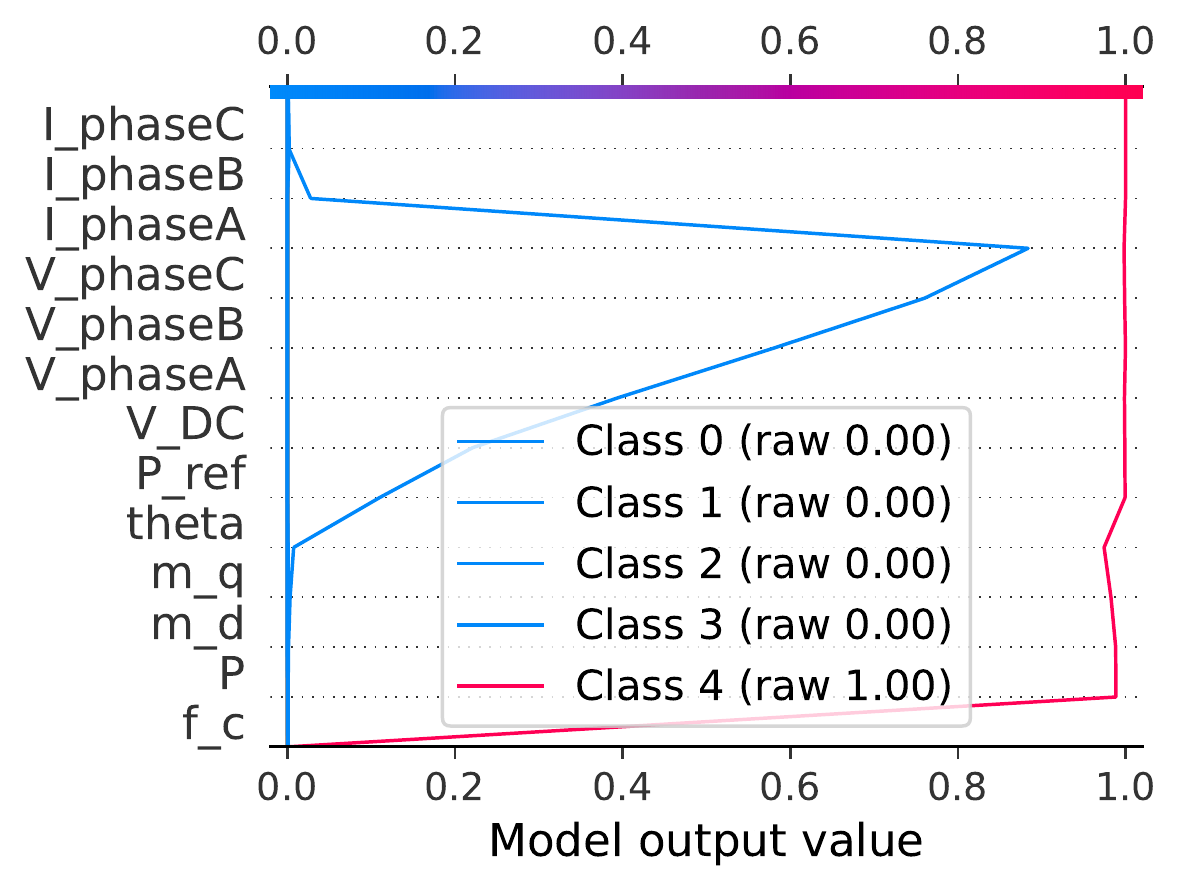}
        \caption{Three-phase grid}
        \label{fig:Three_Phase_Grid_Fault_shap}
    \end{subfigure}%
    \begin{subfigure}{0.4725\linewidth}
        \includegraphics[width=\linewidth]{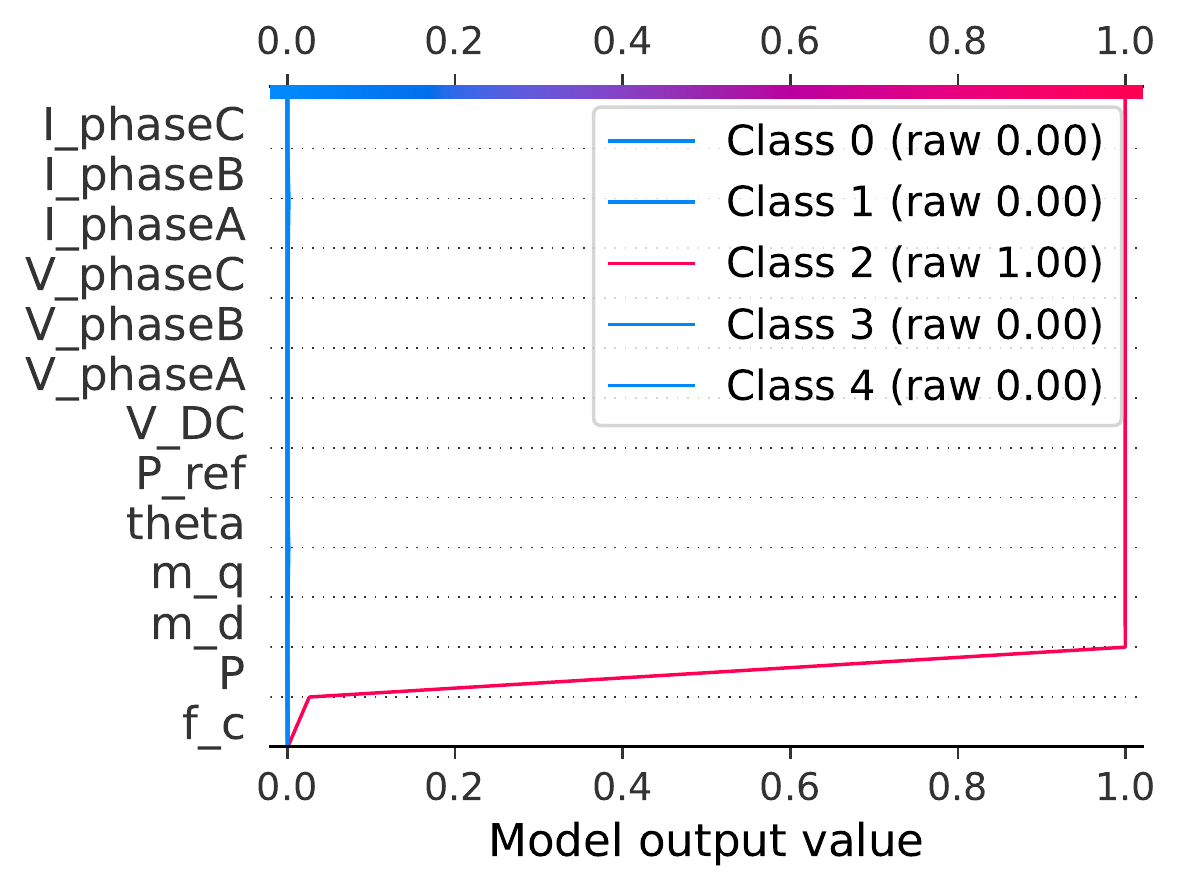}
        \caption{Three-phase sensor}
        \label{fig:Three-Phase_Sensor_Fault_shap}
    \end{subfigure}%
\captionsetup{font=footnotesize}
\caption{LightGBM SHAP value examples.}
\label{fig:explanation}
\end{figure}

\section{Results and Discussion}
%For production purpose, it is possible to upgrade the license or use an open-source solution, e.g., MLFlow~\cite{chen2020developments}.
%To highlight the problem more clearly, 
For performance assessment, the system in Fig.~\ref{fig:sys} is evaluated under two instances of disturbances; in the first case, only a single-class fault is introduced, while in the second case, two consecutive faults are introduced at $t$=$1$ sec, as illustrated in Fig.~\ref{fig:mlr}.
In a limited data setting, system behavior might compromise the digital twin operation, as their dynamic response is highly similar. To address this issue, we first evaluate the baseline methods, i.e., the best H2O AutoML method with the lowest mean per class error in the default AutoML leaderboard, and the traditional model.
Next, we perform hyperparameter fine-tuning on LightGBM and deep learning multilayer perceptron models. The parallel coordinate plots in Fig.~\ref{fig:p_coor_lgbm} illustrate the effect of each parameter on the model fine-tuning, while 
%For partial example results, see parallel coordinate plots showing effect of each parameter on the model fine-tuning in Figs.~\ref{fig:p_coor_lgbm}, and 
the effect of model parameters on focal loss function and accuracy of the LightGBM model are depicted in Figs.~\ref{fig:f_loss_lgbm} and \ref{fig:acc_lgbm}, respectively.

The classification outcomes of LightGBM model with the best parameter combination are summarized in  Table~\ref{lgbm_cls_report}.
Standard classification metrics (i.e., accuracy, precision, recall, F1-score) are evaluated for combined validation and test dataset predictions from the model with the best performance on the training dataset. Higher precision values indicate low false positive rates, while higher recall values indicate low false negative rates. A higher F1-score indicates better detection performance.
We conclude that for the given dataset and task, the model with the highest accuracy is fine-tuned LightGBM with custom multi-class focal loss function.

\begin{figure*}[t!]
	\centering
	\includegraphics[width=\textwidth]{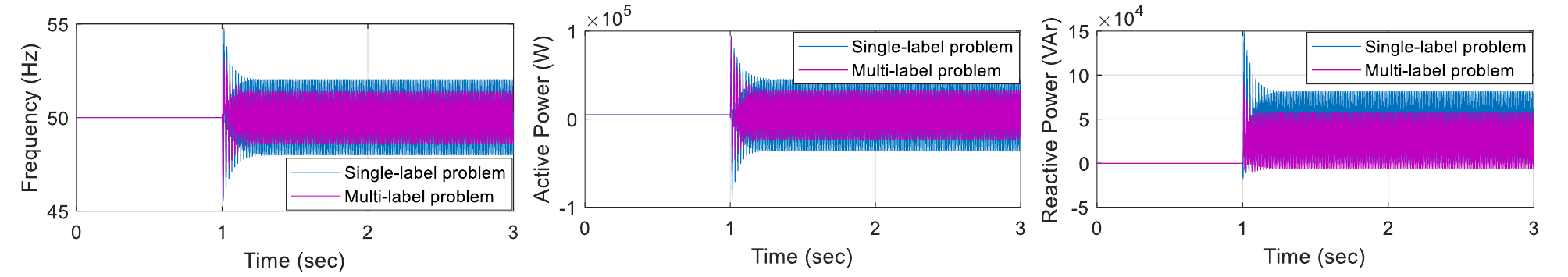}
    \captionsetup{font=footnotesize}
	\caption{Performance of VSG controlled grid-forming converter during two classes: (a) single-label -- Phase A voltage sag of 0.5 pu. is introduced at t=1 sec; (b) multi-label -- Phase A voltage sag and voltage sensor fault in phase B are introduced simultaneously at t=1 sec.}
	\label{fig:mlr}
\end{figure*}

%Individual data analysis components are publicly available in the form of Jupyter notebooks, scripts and accompanying code on the authors GitHub repository. The aforementioned pre-existing platform is implemented in Docker\footnote{Docker [Online]: https://www.docker.com.} containers and is also publicly available.

% \begin{figure}
% \subfloat[\label{fig:experiment_tracking_cropped}Experiment tracking, model versioning and deployment]{\includegraphics[width=0.45\textwidth]{Figures/experiment_tracking.pdf}}
% \end{figure}

\begin{figure*}[t!]
\centering
\begin{subfigure}{\textwidth}
\centering
\includegraphics[width=0.9\textwidth]{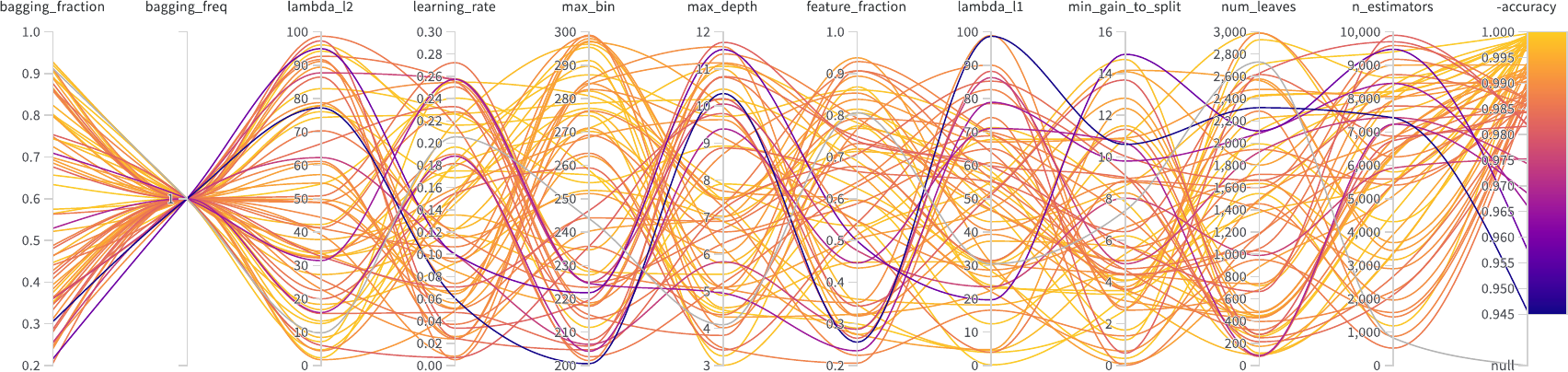}
\caption{\footnotesize Parallel coordinates}
\label{fig:p_coor_lgbm}
\end{subfigure}%
\vspace{0.33cm}
\begin{subfigure}{\textwidth}
\centering
\includegraphics[width=0.5\textwidth]{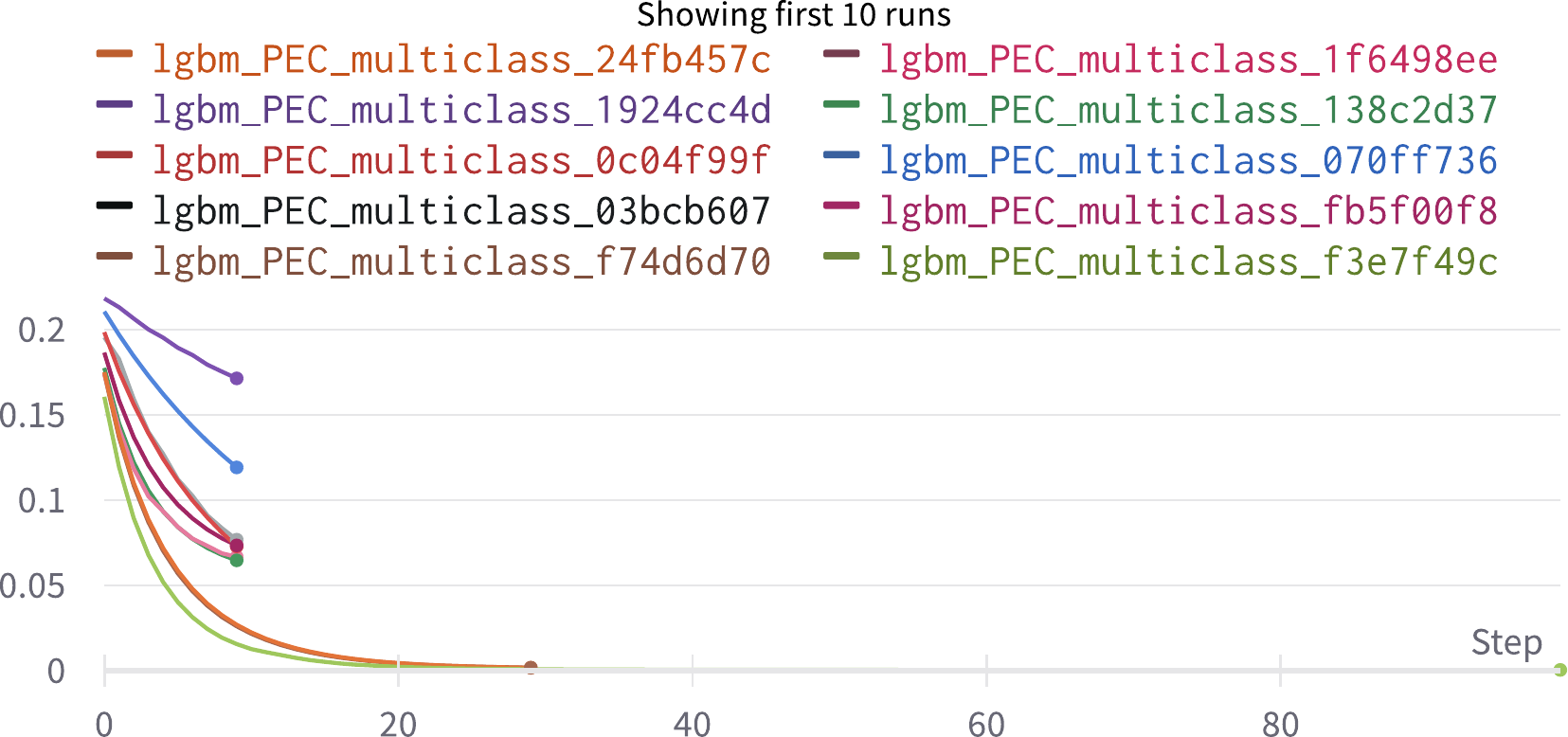}
\caption{\footnotesize Focal loss}
\label{fig:f_loss_lgbm}
\end{subfigure}%
\vspace{0.33cm}
\begin{subfigure}{\textwidth}
\centering
\includegraphics[width=0.5\textwidth]{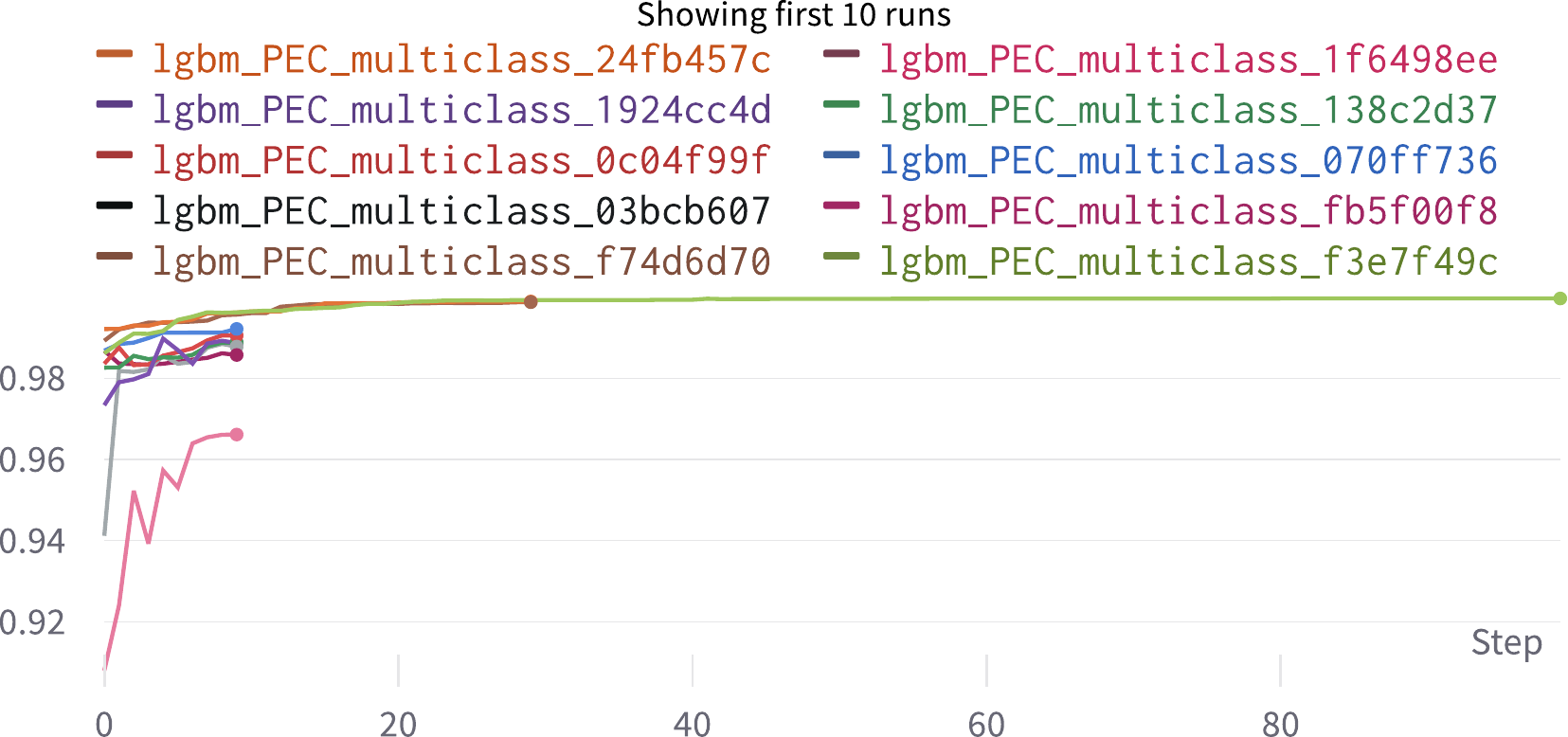}
\caption{\footnotesize Accuracy}
\label{fig:acc_lgbm}
\end{subfigure}%
\captionsetup{font=footnotesize}
\caption{LightGBM training. The parallel coordinates in Fig.~\ref{fig:p_coor_lgbm} reveal the effect of the algorithm parameter combinations on the accuracy of the trained models. Figs.~\ref{fig:f_loss_lgbm} and~\ref{fig:acc_lgbm} demonstrate the training process (steps on horizontal axis), where models with low-performing parameter combinations are stopped early by ASHA scheduler to save resources for high-performing parameter combinations, i.e., those exhibiting low focal loss and high accuracy.}
%with promising metrics results - low focal loss/high accuracy, early in the training process.}
\end{figure*}

\begin{table}[t!]
\centering
\captionsetup{font=footnotesize}
\caption{LightGBM classification report.}
\resizebox{0.48\textwidth}{!}{
\begin{tabular}[b]{|l|c|c|c|c|}
    \hline
	\textbf{Class} & \textbf{Precision} & \textbf{Recall} & \textbf{F1-score} \\ \hline
Normal behavior & 0.99 & 0.99 & 0.99 \\ \hline
Line-to-line fault & 0.88 & 0.72 & 0.79 \\ \hline
Sensor fault & 0.99 & 1.0 & 0.99 \\ \hline
Single-phase voltage sag & 0.99 & 0.99 & 0.99 \\ \hline
Three-phase faults & 0.93 & 0.75 & 0.83 \\ \hline
\textbf{Macro average} & 0.96 & 0.89 & 0.92 \\ \hline
\textbf{Weighted average} & 0.99 & 0.99 & 0.99 \\ \hline\hline
\textbf{Accuracy} & \multicolumn{3}{c|}{0.99}  \\ \hline
\end{tabular}
}
\renewcommand{\arraystretch}{1.3}
%\subfloat[\label{fig:conf_m_lgbm}Confusion matrix]{
%\includegraphics[width=0.45\textwidth]{figures/lgbm_default_confusion_matrix.png}}
\label{lgbm_cls_report}
\end{table}

\section{Conclusions and Future Work}\label{sec:conclusion}
This paper proposes a framework to optimize the operation of digital twins employed for fault diagnosis of grid connected inverters by extending the search space using online data via a Bayesian approach. In summary, we can not only enhance the accuracy of digital twin (designed using limited data) over time, but can also formalize \textit{high confidence} over its decisions. From an implementation perspective, this approach is more practical, as it is difficult to render different classes of fault data for selective decisions. 

In the path forward, we plan to use TinyML solutions to deploy our optimization framework in an experimental testbed, and document its ability to operate as an edge digital twin for power electronics. We also plan to extend our mechanism for online stability assessment to simplify the efforts required in computational modeling and analysis. 
The applicability of the proposed methodology in a horizontal federated learning~\cite{yang2019federated} setup will be also explored, aiming to address use cases of training shared prediction model for multiple power electrical circuits. This approach can be directly applied using proposed approaches for GBMs~\cite{yang2019tradeoff} and deep learning models using federated averaging algorithm~\cite{mcmahan2017communication}.

\section*{Acknowledgment}
This work is partly supported by the FIREMAN project CHIST-ERA-17-BDSI-003 funded by the Spanish national foundation (PCI2019-103780) and Academy of Finland (AoF; n.326270), also AoF via EnergyNet Research Fellowship (n. 321265/328869/352654), and X-SDEN project (n. 349965), and Nordic Energy Research via Next-uGrid project (n. 117766).

\bibliographystyle{IEEEtran}
\bibliography{bibliography}

\end{document}